\documentclass[12pt]{iopart}
\usepackage{iopams}
\usepackage{graphicx}
\usepackage{psfrag}
\usepackage{dcolumn}
\usepackage{latexsym}
\usepackage{amssymb,latexsym,mathrsfs,color}
\usepackage{amsfonts}

\newcommand{\beq}{\begin{equation}}
\newcommand{\eeq}{\end{equation}}
\newcommand{\beqn}{\begin{eqnarray}}
\newcommand{\eeqn}{\end{eqnarray}}
\newcommand{\bearr}{\begin{array}}
\newcommand{\enarr}{\end{array}}

\newcommand{\ra}{\rangle}

\def\bea{\begin{eqnarray}}
\def\eea{\end{eqnarray}}
\def\ba{\begin{array}}
\def\ea{\end{array}}

\def\theta{\alpha}

\def\clr{\textcolor{black}}

\begin{document}
 
\title{Absorbing  Phase Transition in a Four State Predator Prey Model in One Dimension}
\author{Rakesh Chatterjee, P. K. Mohanty and Abhik Basu}
\address{Theoretical Condensed Matter Physics Division,\\ Saha Institute of Nuclear Physics, Kolkata 700064, India.}
\ead{rakesh.chatterjee@saha.ac.in}

\def \rb {\rho^{s}_{_B}}
\def \ra {\rho^{s}_{_A}}
\def \rbt {\rho_{_B}}
\def \rat {\rho_{_A}}

\begin{abstract}
The model of competition between densities of two different species, called predator and prey, is studied on a one dimensional periodic lattice, where each site can be in one of the four states say, empty, or occupied by a single predator, or occupied by a single prey, or by both. Along with the pairwise death of predators and growth of preys, we introduce an interaction where the predators can eat one of the neighboring prey and reproduce a new predator there instantly. The model shows a non-equilibrium phase transition into a unusual absorbing state where predators are absent and the lattice is fully occupied by preys. The critical exponents of the system are found to be different from that of the Directed Percolation universality class and they are robust against addition of explicit diffusion.
\end{abstract}


\maketitle

Absorbing configurations do not have any outgoing rates \cite{AAPTBook}. Once
reached there, the system can not escape from these
configurations. Presence of absorbing configurations in a
phase space raises a possibility that the concerned system may undergo
a non-equilibrium phase transition into absorbing states. The critical behavior
of these absorbing state phase transitions (APT)s \cite{DPBook}
depends on the conservation in dynamics and the symmetry between
absorbing states. It has been conjectured \cite{DPconj} that in absence of any
special symmetry the APT belong to the directed percolation (DP) 
universality class as long as the system has a single absorbing state.
Additional symmetries, like particle-hole symmetry \cite{u9},
conservation of parity \cite{u10}, and symmetry between different absorbing
states \cite{u11} lead to different universalities. Spreading process 
with spatially quenched randomness \cite{qnchd} or with long-term memory \cite{memory} are known to
destroy the critical behavior completely, whereas the 
\clr{long-range interaction leads to continuous variation \cite{longrange} of critical exponents.}
Presence of infinitely many
absorbing states may \cite{inf_yes} or may not \cite{inf_no} belong to
DP universality class.
Again a different critical behavior is observed when the activity
field does not have any special symmetry, but it is coupled to a
conserved density \cite{CTTP}. Recent studies have indicated that DP-critical behavior
is possible, even in presence of an additional conserved field \cite{dd}.
It is not quite clear, what microscopic ingredients can make an APT
belong to the DP class.

The models of directed percolation has been extended to more than one species \cite{twospc}.
Along with the simple DP behavior, a line of first order transition \cite{Dickman:91}
has been observed in $1+1$ dimension when two species compete for survival.
Janssen \cite{Janssen:97} studied coupled DP processes
with bilinear and bidirectional interspecies couplings in the
framework of bosonic field theory, where no other critical
phenomena were found other than the DP.
Hierarchy of unidirectionally coupled DP processes with many species
show multicritical behavior \cite{Tauber:98}.
Coupled percolation processes have been also studied \cite{Cardy:96},
where the absorbing phase become unstable with respect to
an arbitrarily small branching rate even in one dimension.

Predator-prey cellular automaton models \cite{Tome:07} in two dimension show DP universality class. Coupled directed percolation (DP) processes with more than two species
of particles (in one dimension) with different kind of interspecies coupling
have shown DP-type \cite{Park:04} transitions. Lotka-Volterra like models in one dimension
always show coexistence \cite{Tauber:07}, either in form of well mixed states or as irregular
bursts of the predator and prey population. A four state predator prey model \cite{lipowska} in 
one dimension with a restriction that a site can have at best one particle of each kind,
shows an APT to an absorbing (extinct) state which belongs to DP-class.

In this article we study a model of two species, say $A$  
(prey) and $B$ (predator), on a  $(1+1)$-dimensional lattice. Each lattice site  
is either vacant $\O$ or occupied by at best one particle of each kind. 
The preys grow independently as $A\O\to AA$ and the predators die as
$BB\to\O\O$, whereas they interact through a process $BA\to BB$, where 
birth of a new predator occurs instantly along with the death of the prey. 
The system show a line of continuous absorbing state transition different from DP as the 
rates of these processes are tuned.
Unlike other multispecies models, in the absorbing state both preys and predators do not separately extinct, rather predators extinct and preys proliferate to fill up the whole prey branch. 
\begin{figure}[t]
\centering
 \includegraphics[width=8cm,bb=14 14 457 119]{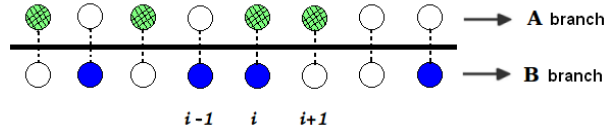}
\caption{Schematic diagram of the 4-state predator-prey (4SPP) model}
\label{fig:model}
\end{figure}
The model is defined on a one dimensional periodic lattice with lattice sites labelled by $i=1,2\dots L$. Each site $i$ can be either vacant  or occupied by a single particle $A$ (prey), or occupied by a single particle $B$ (predator) or by both particles (co-existing $A$ and $B$), thus the model can be treated as 4-state predator-prey (4SPP) model. More than one particle of any kind is not allowed. These hardcore restrictions on individual particles, where co-existence is allowed, 
can be realized  alternatively by  considering two separate branches, one for $A$ and the other for $B$ particles. 
Particles living in one branch can not move to the other branch, as schematically shown in the Fig. \ref{fig:model}. Correspondingly, each site  $i$ is associated with  four states;  $s_A^i = 0,1$   and  $s_B^i = 0,1$,  where $1$ ($0$) denotes the presence (absence) of a particle at site $i$.

On a periodic lattice, these  \clr{particles interact following a random sequential dynamics}  given below. 
\begin{figure}[h]
 \centering
\includegraphics[width=5cm,bb=14 14 330 271]{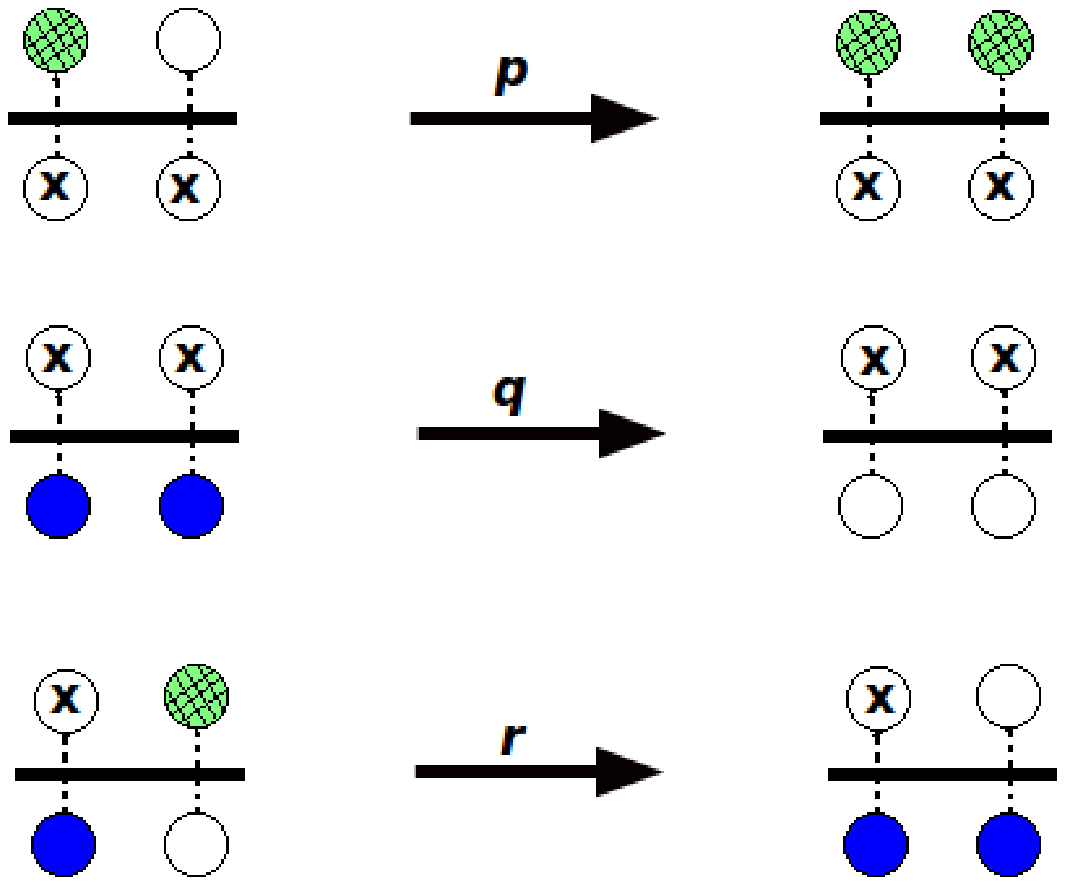}
\end{figure}
The prey ($A$) can grow on their own branch with rate $p$ independent of the predators ($B$). 
Again, two neighboring  predators ($B$) die simultaneously with rate $q$ independent of $A$ due to their own crowding or overpopulation. Here $X$ in $A$-branch ($B$-branch) corresponds to an arbitrary state of $A$ ($B$) particles i.e. presence or absence. These two species interact with rate $r$ as follows; when a predator $B$ at site $i$ meets a prey $A$ as the right neighbor $i+1$, with site $i+1$ is not already occupied by a predator $B$, then the predator $B$ eats the prey $A$ and simultaneously reproduces another predator $B$. \clr{Note, that 
the dynamics is spatially asymmetric as both species grow only in one direction (here, right), and it does not include  explicit diffusion of species.} Effect on addition of symmetry and diffusion are briefly 
discussed towards the end. 

Let the steady state density of $A$ and $B$ particles be $\ra$ and $\rb$ respectively.  
Clearly in absence of predators ($\rb$) the prey density $\ra=1$ as the preys do not  
have an independent death process.
Again, note that the dynamics do not allow $A$ particles to be eaten when they co-exist with $B$ on the same site.  
This indicates that these $A$s can only die after their coexisting $B$s die, which can occur with rate 
$q$ if there is a $B$ particle present or created at the immediate left neighbor. 
Thus, for reasonably small death rate of predators, $\ra$ is expected to have a small value 
($\ra<1$) when $\rb\simeq 0$ and then it 
increases along with $\rb;$ so, the prey density $\ra$ can never vanish. The predator density $\rb$ 
can, however, become zero by repetitive death process. The isolated  $B$s  wait until the  prey invades 
their neighboring site and then they subsequently eat and reproduce with  rate $r$ and die with  rate $q$. So, along with the {\it coexisting phase} \footnote{It has been predicted earlier
\cite{lipowska} that, in absence of site restriction, the  predator and the prey system in 1D always remain in the 
co-existing phase.} where 
both $\ra$ and  $\rb$  are non-zero, we have another phase where $\ra=1$  and  $\rb=0.$  Clearly, the 
later phase $(\ra=1, \rb=0)$ is absorbing as once all the predators die, even then the single surviving 
prey can lead to proliferation of the prey population in the whole lattice. 
Of course, as argued earlier, $\ra$ can not vanish, and the other possible absorbing state $(\ra=0, \rb=0)$ which requires simultaneous death of all predators and preys, is not dynamically accessible. Thus 
the 4SPP model can undergo an absorbing state phase transition by tuning the different rates with   
$\rb$ as the order parameter. Our aim here is to study this critical behavior in details.

We have used the standard Monte Carlo methods to study the critical behavior of this model. From an initial arbitrary configuration, where  each branch $A$ and $B$ are filled by arbitrary number of respective particles, the system is allowed to evolve according to the random sequential update following the Monte Carlo dynamics of 4SPP model. We have studied the system with different values of the rates $p$, $q$ and $r$ with system size $L=10^3$. For illustration, we have fixed two of the reaction rates, say $q=0.02$, $r=0.9$, and vary $p$ as the the control parameter in the following simulations.

\begin{figure}[h]
\centering
\includegraphics[width=5cm,height=4.1cm,bb=50 50 410 302]{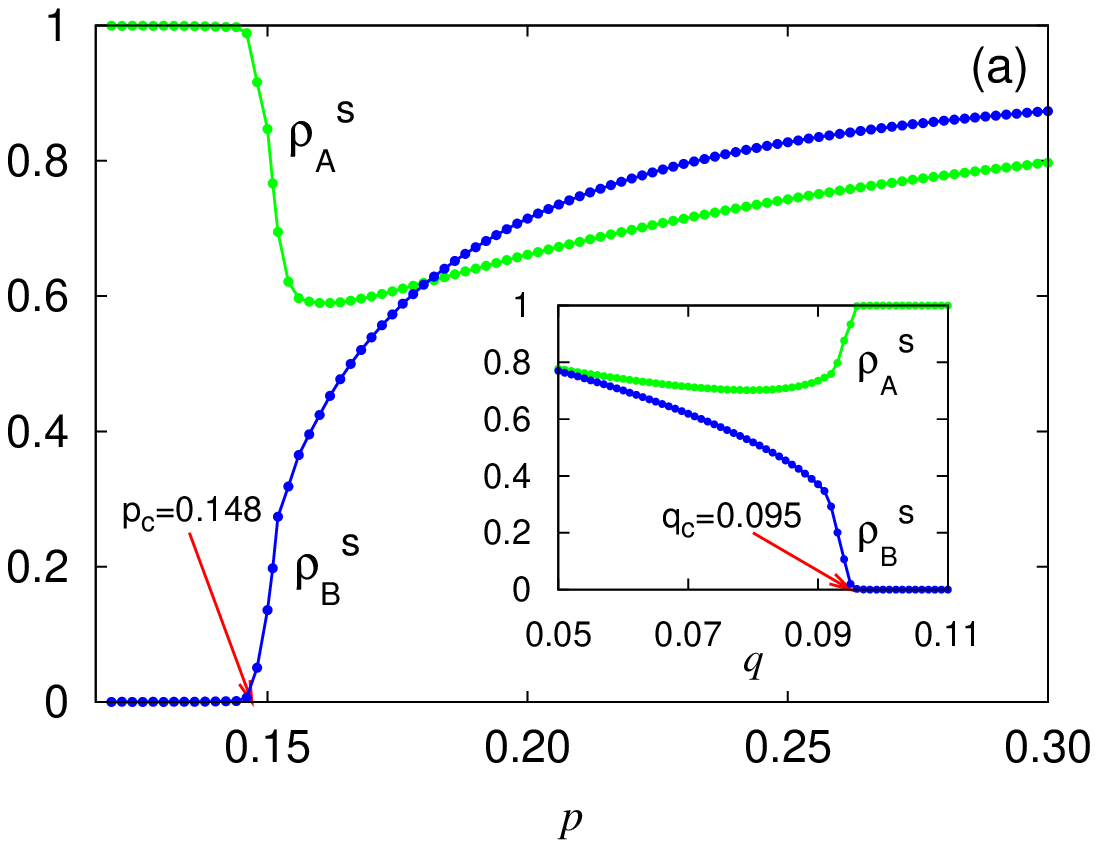}
\includegraphics[width=5cm,height=4cm,bb=50 50 410 302]{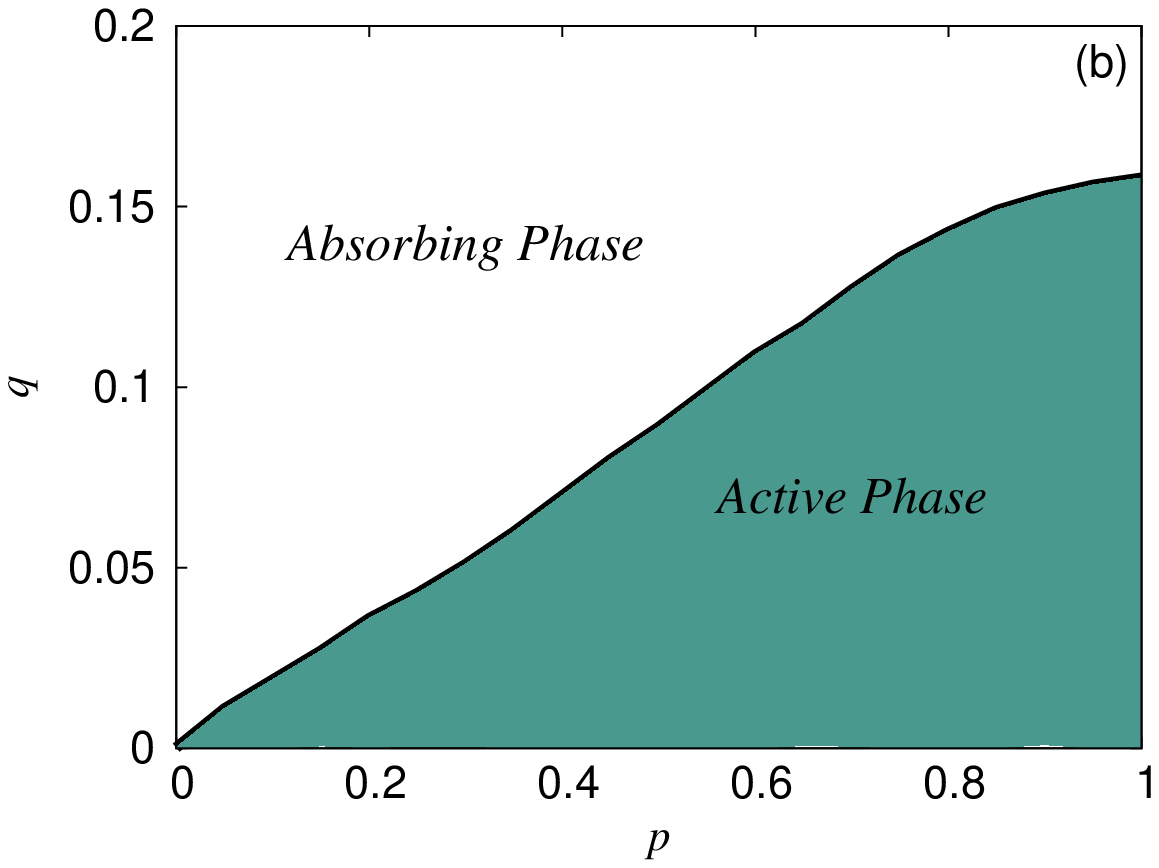}
\caption{(a) Average steady state density $\ra, \rb $ is plotted against the control parameter $p$ for fixed value of $q=0.02, r=0.9$ and $L=10^3$, critical point is indicated at $p_c=0.148(4)$ showing absorbing to active phase transition. Both densities are again plotted against $q$ keeping $p=0.55$, $r=0.9$ fixed; critical point is $q_c=0.095(6)$ showing active to absorbing phase transition (see inset).
(b) \clr{Phase diagram in the ($p,q$) plane for $r=0.9$, showing active and absorbing phases.}}
\label{fig:phase}
\end{figure}

For $p<p_c$ the average $B$ density $\rb$ decreases continuously until the system reaches a state with no predators ($\rb=0$) and ultimately the whole prey lattice branch is filled with preys. Once reached in this absorbing state $(\ra=1, \rb=0)$, the system remains there forever. While, for $p>p_c$ the average density of $B$ particles $\rb$ saturates to a nonzero value along with $\ra$ which also takes a value smaller than unity. In Fig. \ref{fig:phase}(a) we have plotted $\ra$ and $\rb$  as a function of $p$ for a system size $L=10^3.$  As expected, the density of the preys $\ra$ never vanish and they proliferates in the whole lattice (with $\ra=1$) in the absorbing state where predators are absent. Thus the system undergoes an absorbing state phase transition as the birth rate of preys $p$ crosses a critical threshold $p_c=0.148(4).$


\begin{figure}[b]
\centering
 \includegraphics[width=5cm,height=4.1cm,bb=50 50 410 302]{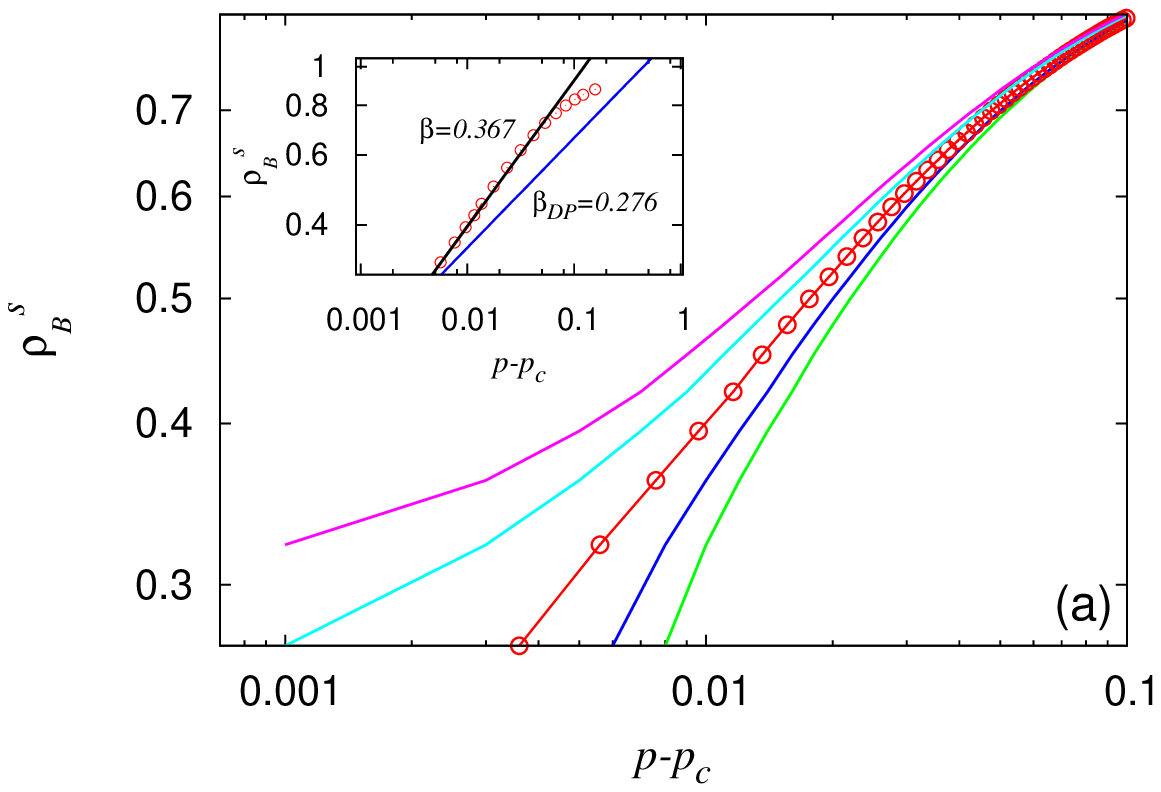}\includegraphics[width=5cm,height=4.1cm,bb=50 50 410 302]{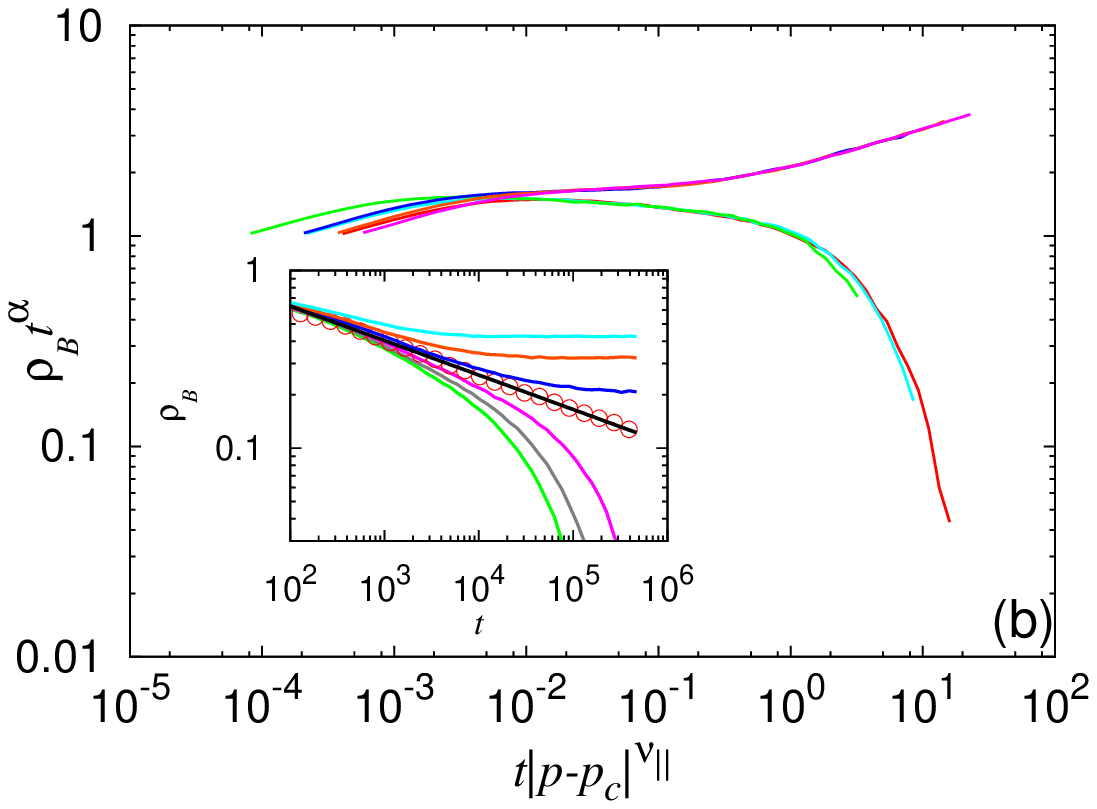}
\caption{(a) Average predator density ($\rb$) is plotted with fixed value of $q=0.02$,$r=0.9$; with different choices of $p_c=0.153,0.151,0.148,0.146,0.144$ shown from top to bottom. The correct choice of $p_c=0.148(4)$ and corresponding slope in logarithmic plot gives critical exponent $\beta=0.367(7)$. (b) Time evolution of $\rbt$ with $q=0.02$,$r=0.9$. Below the critical point for $p=0.142, 0.144, 0.146$ density $\rbt$ eventually extincts and above the critical point $p=0.150, 0.154, 0.160$ density $\rbt$ saturates. At the critical point $p=0.148(4)$, $\rbt(t) \sim t^{-\theta}$ gives the critical exponent $\theta=0.194(4)$ and the data collapse gives $\nu_\shortparallel=1.8(1)$.}
\label{fig:beta_nu}
\end{figure}

For $p>p_c$, the order parameter $\rb$ shows power law behavior with the distance from criticality,
\begin{equation} 
\rb \sim (p-p_c)^\beta,
\label{eq:beta}
\end{equation}  
when $p$ approaches $p_c.$ This Eq. (\ref{eq:beta}) can be used to estimate $p_c$ and $\beta$. As shown 
in  Fig. \ref{fig:beta_nu}(a), $\rb$ versus $(p-p_c)$ is linear in log scale, for the correct choice of $p_c=0.148(4)$; the corresponding slope $\beta=0.367(7)$ gives the estimated value of the order parameter exponent.
 
One can obtain few other critical exponents from the decay of the order parameter $\rbt(t)$ from an 
initial configuration with large number of predators. Clearly, in the vicinity of critical point $\rbt$ is a 
function of time $t$ and the temporal correlation length $\xi_\shortparallel$ which vary as
$\xi_\shortparallel \sim |p-p_c|^{-\nu_\shortparallel}$. Again, after an initial decay  
$\rbt(t)\sim t^{-\theta}$ the predator density in the active phase approaches the steady state value $\rb$ in  the $t\to\infty$ limit. So $\rbt$  must scale as,
\begin{equation}
\rbt(t, p) = t^{-\theta} {\cal F}(t |p-p_c|^{\nu_\shortparallel}).
\label{eq:rhot_eps}
\end{equation}
Thus, one expects that $\rbt(t)$ for different values of $p$ (shown in the Fig. \ref{fig:beta_nu}(b)) collapsed into a single scaling function ${\cal F}$, when $\rbt t^{\theta}$ is plotted against $t |p-p_c|^{\nu_\shortparallel}$. The main figure here shows the data collapse when we choose $\theta=0.194(4)$,
and $\nu_\shortparallel=1.8(1).$ Since at the critical point $\rbt(t, p_c)=  t^{-\theta} {\cal F}(0)$, one can obtain both $p_c$ and $\theta$ directly from the log scale plot of $\rbt$ versus $t$ which is linear (as shown in the inset of Fig. \ref{fig:beta_nu}(b)). The resulting $p_c$ and $\theta$ are consistent with those obtained from the data collapse.
Again, in the upper critical regime, $\rb$ vanishes as $|p-p_c|^\beta$, in the $t\to \infty$ limit. 
This can happen only when the off-critical scaling function ${\cal F}(x)\sim x^{\beta/\nu_\shortparallel}$;
thus $$\theta = \beta/\nu_\shortparallel$$
Since all three exponents $\beta$, $\theta$ and $\nu_\shortparallel$ are calculated independently, one can check if the above scaling relation holds. In this case it holds to a great accuracy for the values
of $\beta$, $\theta$ and $\nu_\shortparallel$ calculated here.


\begin{figure}[b]
\centering
\includegraphics[width=5cm,height=4.1cm,bb=50 50 410 302]{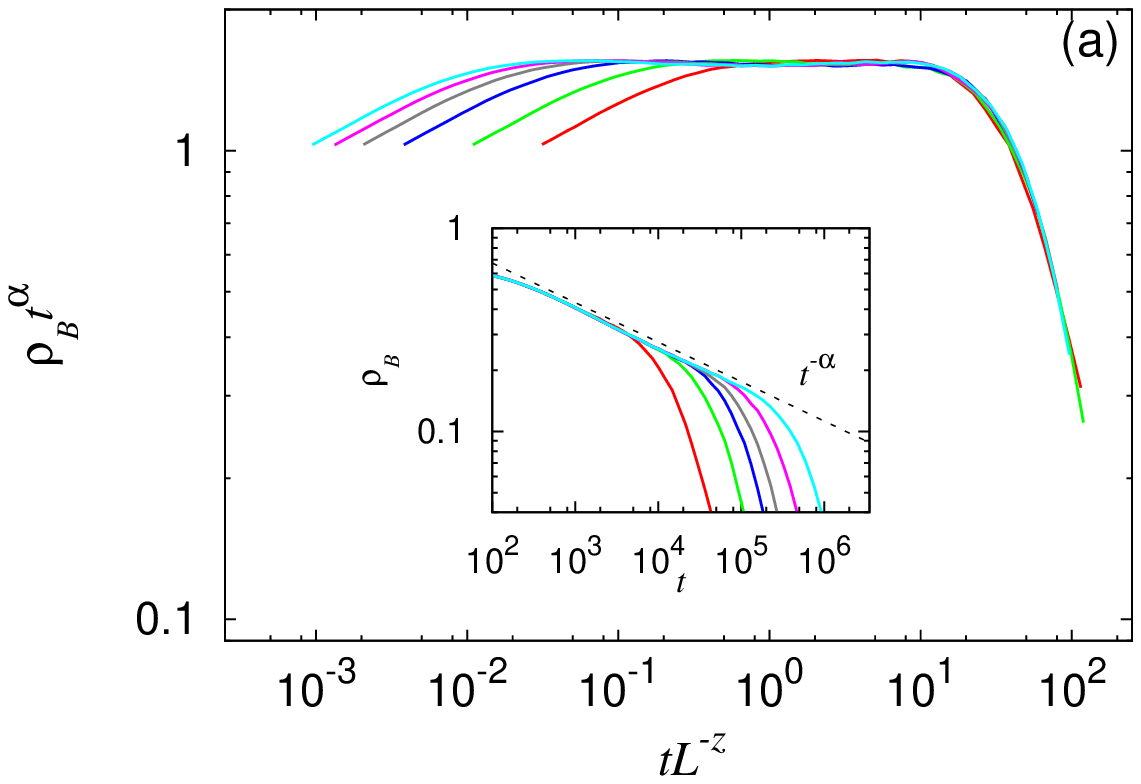}\includegraphics[width=5cm,height=4.1cm,bb=50 50 410 302]{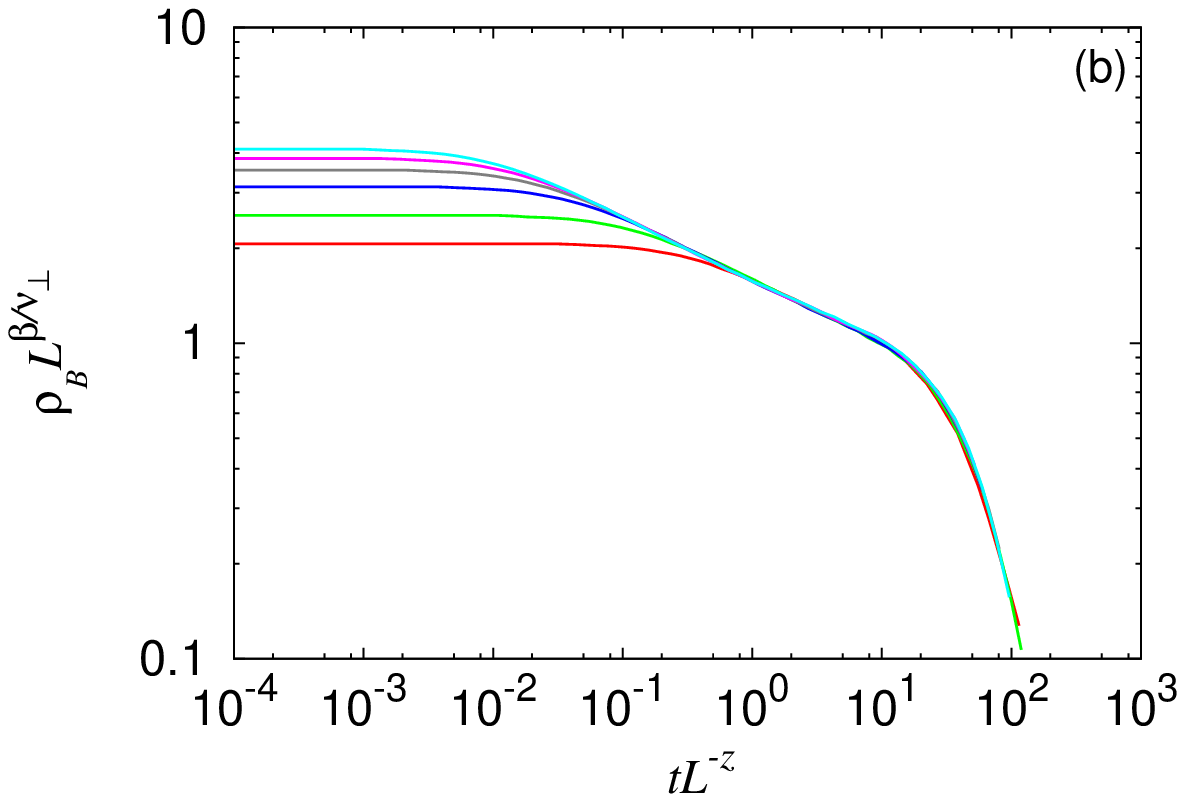}
\caption{(a) Finite size scaling function $\rbt t^\theta$ plotted against the scaled variable 
$tL^{-z}$ for $L=50,100,200,300,400,500$ (bottom to top in figures) with $p=0.148(4)$, estimates $z=1.52(0)$. (b) The same data could be collapsed according to Eq. (\ref{eq:rhot_L1}) by choosing  $\beta/\nu_\perp=0.30,$ which gives an estimate $\nu_\perp=1.2(2).$ }
\label{fig:zed_nu}
\end{figure}

Now we turn our attention to the finite size scaling of $\rbt$ at the critical point.
Again, the system  of length $L$ with a high density of predators $\rbt(t,L)$ decays as       
$t^{-\theta}$, indicating a scaling form
\begin{equation}
\rbt(t, L) = t^{-\theta} {\cal G}(t/L^z),
\label{eq:rhot_L}
\end{equation}
where $z$ is the dynamic critical exponent. Thus, one expects $\rbt$ for different values of $L$ to be collapsed to a single function when plotted against $t/L^z$. This is described in Fig. \ref{fig:zed_nu}(a).
The inset there shows variation of $\rbt(t)$ for different system size $L=50,100,200,300,400,500$, which were made to collapse to a single function using $\theta=0.194(4)$ and $z=1.52(0)$.  
From the scaling relation $$z= \nu_\shortparallel/\nu_\perp,$$  one  expects that  $\nu_\perp= 1.19.$ This can be verified from the modified scaling relation Eq. (\ref{eq:rhot_L}). Since  $ z \theta = \beta/ \nu_\perp$, we have  
\begin{equation}
\rbt(t, L) = L^{-\beta/ \nu_\perp} \tilde {\cal G}(t/L^z),
\label{eq:rhot_L1}
\end{equation}
where $ \tilde {\cal G}(x) = x  {\cal G}(x).$ In Fig. \ref{fig:zed_nu}(b) we have plotted $\rbt L^{\beta/ \nu_\perp}$ as a function of $t/L^z$ and found that the data for system size $L=50,100,200,300,400,$ and $500$ 
could be collapsed into a single curve by choosing  $\beta/ \nu_\perp =0.30$. This gives  us $\nu_\perp=1.2(2),$
which is consistent with  the earlier estimation. 

The critical exponents of the model are summarized in Table-\ref{tab1} along with the the critical exponents
of directed percolation universality class. Clearly the exponents, particularly $\beta$ and $\theta$, are very different from those of DP-class, indicating that the 4SPP  model belongs to a different universality class. 
We have studied the model by varying other rates also. 

\begin{table}[h]
\begin{center}
\begin{tabular}{|c|c|c|c|c|c|}
\hline
        & $\beta$         & $\theta$        & $\nu_\shortparallel$ & $z$    & $\nu_\perp$   \\
\hline
DP      & $0.276$         & $0.159$         & $1.733$       & $1.580$       & $1.096$       \\
\hline
4SPP    & $0.367(7)$      & $0.194(4)$      & $1.8(1)$      & $1.52(0)$      & $1.2(2)$        \\
\hline
\end{tabular}
\caption{\label{tab1} Comparison of the critical exponents between the 4SPP model with the DP universality class.}
\end{center}
\end{table}
For example, one can take $q$ as the control parameter, keeping $p$ and $r$ fixed. The inset of Fig. \ref{fig:phase}(a) shows the variation of $\ra$ and $\rb$ as a function of $q$ for fixed $p=0.55$ and $r=0.9$.
Evidently, the order parameter $\rb$ vanishes continuously as $q$ crosses the threshold value $q_c=0.095(6).$  
The critical exponents obtained in this case was found to be consistent with Table-\ref{tab1}. Thus, for any 
fixed values of $r$, one expects a line of criticality in the $(p,q)$-plane, which is shown in Fig. \ref{fig:phase}(b). 

\begin{figure}[h]
 \centering
\includegraphics[width=4.5cm,height=4.5cm]{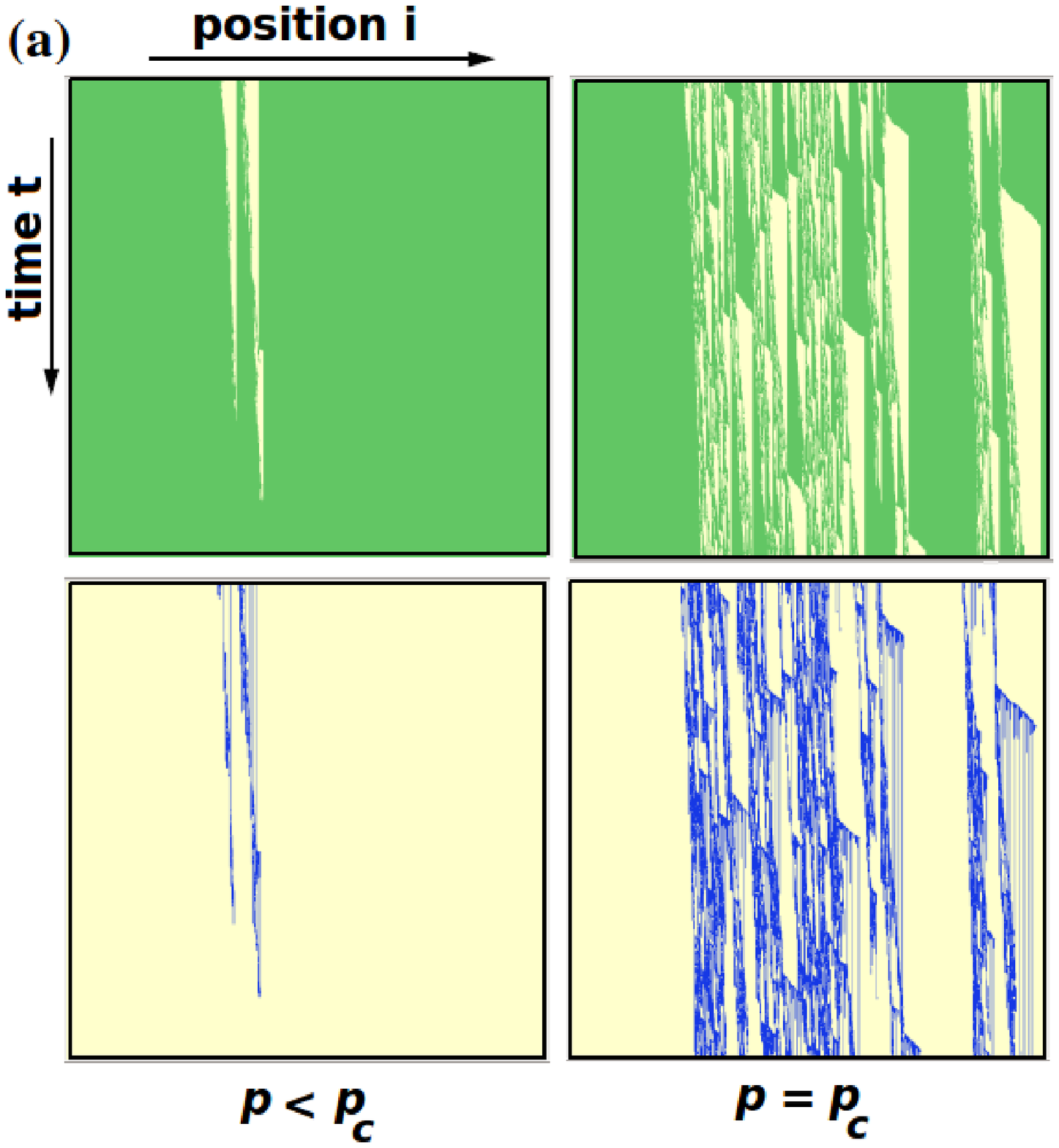}\includegraphics[width=4.5 cm,height=4.5cm]{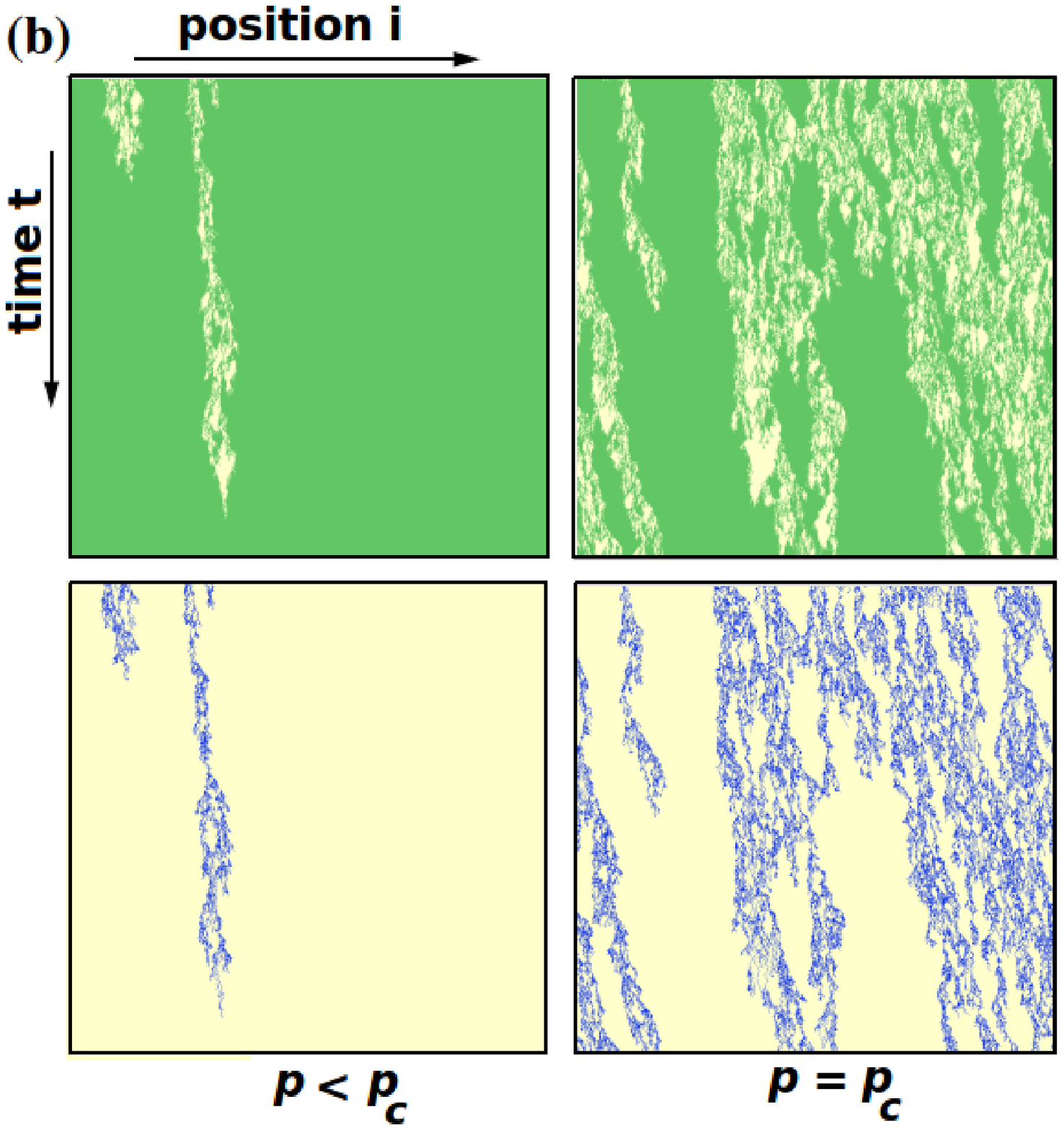}
\caption{Time evolution of the 4SPP model for $10^3$ time steps with prey $A$ (green) and predator $B$ (blue) with system size $L=10^3$ for the regime $p<p_c$ and at $p=p_c$: (a) with asymmetric rules  (no diffusion),
(b) with asymmetric rules and \textit{explicit diffusion}.}
\label{fig:space_time}
\end{figure}

That, the critical behavior of the absorbing phase transition observed in the 4SPP model is different from DP, can be visualized from the growth of clusters. The space-time diagram, starting from  an arbitrary initial configuration is shown in Fig. \ref{fig:space_time}(a), where the occupancy of species $A$ and $B$ are represented separately in the upper and lower part respectively. The evolution of clusters are visibly different from that of $1+1$ dimensional directed percolation model. Since the prey species can grow independently, the space is always filled locally by preys where predators are absent. Again, regular {\it striped} structures  appear in these figures as the species do not diffuse. A natural question would be whether diffusion can drive the system to have an absorbing state transition belongs to DP universality class or not.


In the following we introduce diffusion of both the species \textit{explicitly} in this model. Along with the usual dynamics of the 4SPP model described earlier, both $A$ and $B$ particles are now allowed to move to the neighboring available vacant space in their respective branches. In this case, it turns out that the clusters evolve more like the DP model (see the space-time diagram in Fig. \ref{fig:space_time}(b)). However, the detailed study of the absorbing phase transition reveals that the critical exponents are same as given in Table-\ref{tab1}.

\begin{figure}[h]
\centering
 \includegraphics[width=5cm,height=4.1cm, bb=50 50 410 302]{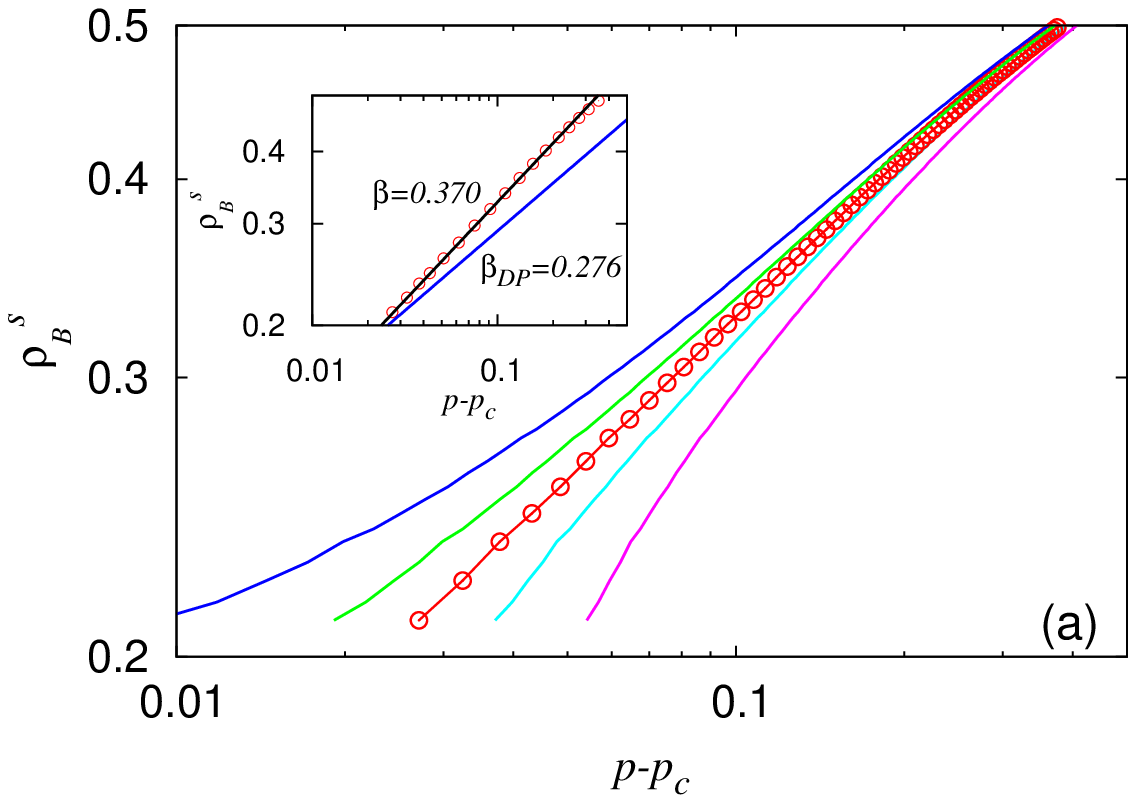}
\includegraphics[width=5cm,height=4.1cm, bb=50 50 410 302]{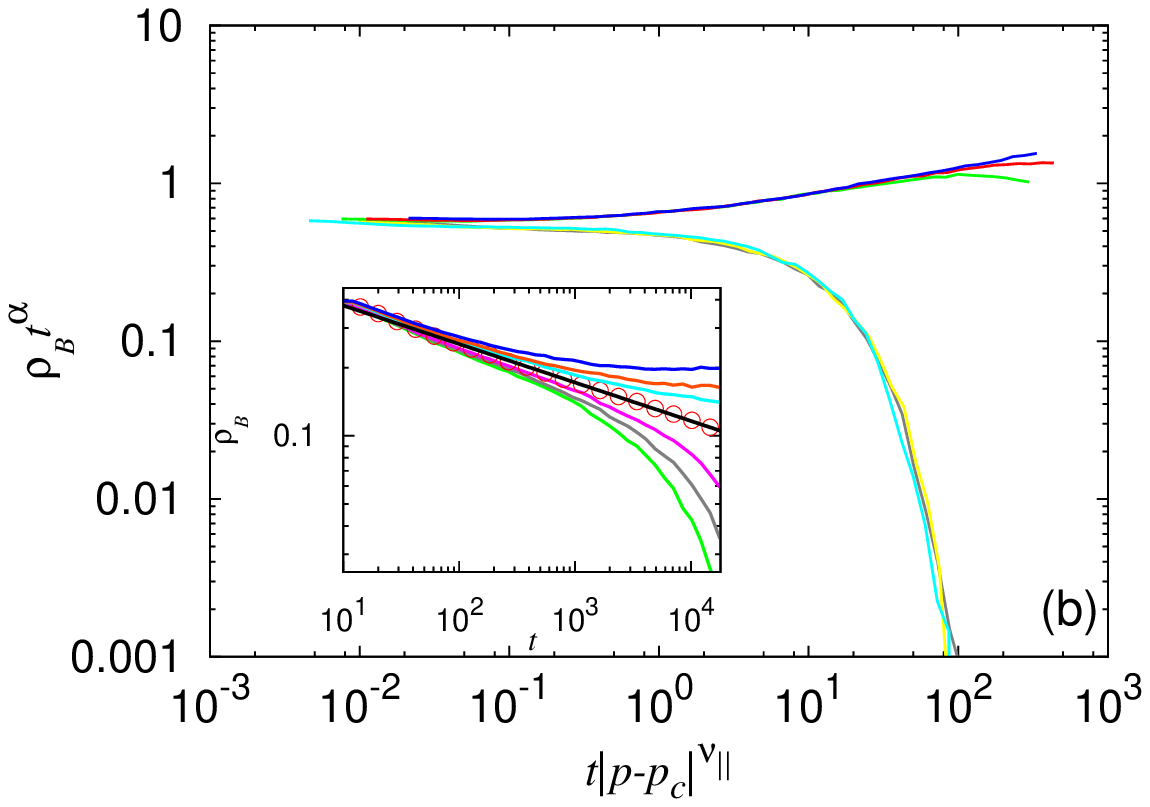}
\caption{(a) Average density of predators ($\rb$) with $q=0.2$, $r=0.9$ with {\it explicit diffusion}: absorbing to active state phase transition at $p=0.323(5)$ and the critical exponent $\beta = 0.370(9)$ (see inset). 
(b) Above the critical point for $p=0.330, 0.335, 0.345$ the density $\rb$ saturates and below the critical point for $p=0.305, 0.310, 0.315$ density $\rb$ eventually extincts. At the critical point $p=0.323(5)$, $\rbt(t) \sim t^{-\theta}$ gives the critical exponent $\theta=0.190(5)$ and $\nu_\shortparallel=1.7(5)$. } 
\label{fig:beta_nu_dfsm}
\end{figure}

In presence of \textit{explicit diffusion}, we choose to study the system with fixed rates $q=0.2$,$r=0.9.$ 
Monte-carlo simulations show that the predator density $\rb$ vanishes continuously as
$p$ is decreases below a critical threshold  $p_c =0.323(5).$ As described in Fig. \ref{fig:beta_nu_dfsm}(a), 
near the critical point, $\rb \sim (p-p_c)^\beta$ with $\beta= 0.370(9).$ Again, starting from a large number of predators, the density $\rbt(t)$ decays to its stationary value $\rb$  which is nonzero only in the  
upper-critical region $p>p_c.$ We find that $\rbt(t)$ for different values of $p$ could be merged to an unique scaling function which satisfy Eq. (\ref{eq:rhot_eps}) by choosing $\theta = 0.190(5)$ and $\nu_\shortparallel=1.7(5).$
This data collapse is described in Fig. \ref{fig:beta_nu_dfsm}(b). The critical exponents $\beta,\theta$ and $\nu_\shortparallel$ are more or less consistent with those listed in Table-\ref{tab1}. It is not surprising that 
addition of explicit diffusion did not alter the universal behavior. In fact, though slow, effective 
diffusion of predators was already occurring in the 4SPP model through the rates $q$ and $r$. 

\clr{Some comments are in order here. The fact that the  absorbing state phase transition  in 4SPP model
is different from that of DP can be seen  from the dynamical mean-field analysis (ignoring spatial density 
correlations). The mean-field 
densities $\rat^i= \langle s_A^i\rangle$  and $\rbt^i= \langle s_B^i\rangle$, in the continuum limit, 
evolve as, 
\begin{eqnarray}
\frac{\partial \rat}{\partial t} &=&  p \rat (1-\rat) - r \rat \rbt - v_A \frac{\partial \rat}{\partial x} + D_A \frac{\partial^2 \rat}{\partial x^2}
\label{eq:MFa} 
\end{eqnarray}
\begin{eqnarray}
\frac{\partial \rbt}{\partial t} &=&  - 2q\rbt^2 + r \rat \rbt - v_B \frac{\partial \rbt}{\partial x} + D_B \frac{\partial^2 \rbt}{\partial x^2},
\label{eq:MFb} 
\end{eqnarray}
where $v_{A,B}$ denote the velocities of respective species which appear due to asymmetric dynamics, 
and $D_{A,B}$   are the  coefficients of diffusion. The first term in Eq. (\ref{eq:MFa}) captures the  
growth of a prey restricted by hardcore interactions and the  second term there corresponds to the 
interaction between two species. Again, the first term in Eq. (\ref{eq:MFb}) represents the simultaneous death 
of two predators. Evidently, these mean-field equations have two fixed points: the unstable one  
$(\rat^*=0, \rbt^*=0)$ and  the stable one $(\rat^*=\frac{2pq}{2pq +r^2},  \rbt^*=\frac{pr}{2pq +r^2}).$  
However, when $\rbt \to 0$ faster than $\rat$, one can get another fixed point $(\rat^*=1,  \rbt^*=0)$ 
from Eq. (\ref{eq:MFa}).
This unusual absorbing state $(\ra=1,\rb=0)$, as discussed earlier, raises a possibility that corresponding 
absorbing state phase transition  can be different from the usual APT to  $(\rat^*=0,  \rbt^*=0).$   
Clearly, for $\rat=1$,  the mean-field equation for the predator density $\rho_B$ (from Eq. (\ref{eq:MFb})) is identical to that of the DP. Thus, in higher dimension (larger than the critical dimension) one expects that  
the 4SPP  model results in the  same  mean-field critical exponents ($ \beta^{MF}=1=\nu_{\shortparallel}^{MF}, 
z^{MF}=2$) as  that of DP. Another possible reason for  the new universality class is the 
asymmetric dynamics, which generates density dependent velocity terms $v_{A,B}$. Note that the absorbing state 
phase transition in the asymmetric contact process \cite{ACP} belongs to the DP-class,  
whereas asymmetric updating is a relevant perturbation to models with extremal dynamics \cite{Dickman}. Detailed 
study of the 4SPP model with symmetric dynamics will be reported elsewhere. }

In summary, we introduce a two species (predator $B$ and prey $A$) model in one dimension where each lattice site is either vacant or occupied by, a single predator, a single prey or both. More than one predator or prey are not allowed at any site. The preys are allowed to grow independent of the predators, whereas two predators, if present at  the neighboring sites, die simultaneously. The species interact through a dynamics where the predator produces an offspring by eating a prey from its neighbor. When the predator density $\rb=0$, even a single prey can invade the whole lattice by its independent birth process. Thus, the system has an unusual absorbing state $(\ra=1,\rb=0);$ the other absorbing state  $(\ra=0,\rb=0)$ is not dynamically accessible. Using dynamical Monte-Carlo simulation  we show that the system shows an absorbing state phase transition, as the birth rate of prey $p$ is increased  beyond a critical value $p_c$ keeping death rate of predator $q$ fixed. For a fixed $r$, the line of criticality ($p_c$ as a function of $q$) is governed by a universality class different from the most generic one, namely directed percolation. This critical behavior is found to be robust against addition of explicit diffusion. Note that the dynamical rules of the model is different from other four species predator prey models studied earlier \cite{lipowska} in a way that the predators in 4SPP model can not eat the prey at the same site.
We believe that, non-equilibrium phase transition to a 
unusual and unique absorbing state, may result in different universality class. Further study in this direction could clarify this issue.

\end{document}